\documentclass[letterpaper, 10 pt, conference]{ieeeconf}  
\IEEEoverridecommandlockouts                                                                 
\usepackage{multirow}
\usepackage{booktabs}
\usepackage{array}
\usepackage{amsmath,amssymb,amsfonts}
\usepackage{algorithmic}
\usepackage{graphicx}
\usepackage{textcomp}
\usepackage{xcolor}
\usepackage{verbatim}
\def\BibTeX{{\rm B\kern-.05em{\sc i\kern-.025em b}\kern-.08em
    T\kern-.1667em\lower.7ex\hbox{E}\kern-.125emX}}
\usepackage{comment}
\usepackage{nohyperref}
\definecolor{matplotlib0}{HTML}{1f77b4}
\definecolor{matplotlib1}{HTML}{d62728}
\definecolor{matplotlib2}{HTML}{2ca02c}
\definecolor{matplotlib3}{HTML}{ff7f0e}
\definecolor{matplotlib4}{HTML}{9467bd}
\definecolor{matplotlib5}{HTML}{8c564b}
\definecolor{matplotlib6}{HTML}{e377c2}
\definecolor{matplotlib7}{HTML}{7f7f7f}
\definecolor{matplotlib8}{HTML}{bcbd22}
\definecolor{matplotlib9}{HTML}{17becf}

\usepackage{mathtools}

\usepackage{booktabs}
\usepackage{array}
\usepackage{multirow}
\usepackage{colortbl}
\usepackage{tablefootnote}
\usepackage{threeparttable}
\usepackage[acronym, style=super, nonumberlist]{glossaries}

\usepackage{pgfplots}
\definecolor{color0}{rgb}{0.12156862745098,0.466666666666667,0.705882352941177} 
\definecolor{color1}{rgb}{1,0.498039215686275,0.0549019607843137}
\definecolor{color2}{rgb}{0.172549019607843,0.627450980392157,0.172549019607843} 
\definecolor{color3}{rgb}{0.83921568627451,0.152941176470588,0.156862745098039} 
\definecolor{color4}{rgb}{0.580392156862745,0.403921568627451,0.741176470588235}
\definecolor{colorblue}{rgb}{0.12156862745098,0.466666666666667,0.705882352941177} 
\definecolor{colorgreen}{rgb}{0.172549019607843,0.627450980392157,0.172549019607843} 
\definecolor{colorred}{rgb}{0.83921568627451,0.152941176470588,0.156862745098039} 
\definecolor{colorblack}{rgb}{0,0,0} 
\definecolor{colororange}{rgb}{1,0.56,0} 
\usepgfplotslibrary{fillbetween}
\usepgfplotslibrary{colormaps}
\pgfplotsset{compat=1.16}

\pgfplotscreateplotcyclelist{matplotlib}{
  {matplotlib0},
  {matplotlib1},
  {matplotlib2},
  {matplotlib3},
  {matplotlib4},
  {matplotlib5},
  {matplotlib6},
  {matplotlib7},
  {matplotlib8},
  {matplotlib9}
}

\pgfplotsset{every axis/.append style={
    cycle list name=matplotlib
}}

\usepackage{listings}

\definecolor{code_default}{HTML}{000000}
\definecolor{code_keyword}{HTML}{AC4142}
\definecolor{code_identifier}{HTML}{D28445}

\newacronym{bp}{BP}{Blood Pressure}
\newacronym{sbp}{SBP}{Systolic Blood Pressure}
\newacronym{dbp}{DBP}{Diastolic Blood Pressure}
\newacronym{cvd}{CVD}{Cardiovascular Disease}
\newacronym{ecg}{ECG}{Electrocardiogram}
\newacronym{ppg}{PPG}{Photoplethysmogram}
\newacronym{eeg}{EEG}{Electroencephalogram}
\newacronym{int8}{INT8}{8-bit Integer}
\newacronym{mae}{MAE}{Mean Absolute Error}
\newacronym{bhs}{BHS}{British Hypertension Society}
\newacronym{aami}{AAMI}{Association for the Advancement of Medical Instrumentation}
\newacronym{rnn}{RNN}{Recurrent Neural Network}
\newacronym{cnn}{CNN}{Convolutional Neural Network}
\newacronym{abp}{ABP}{Arterial Blood Pressure}
\newacronym{soa}{SOA}{State-of-the-art}
\makeatletter
\def\ps@IEEEtitlepagestyle{%
  \def\@oddfoot{\mycopyrightnotice}%
  \def\@evenfoot{}%
  \def\@oddhead{}
  \def\@evenhead{}%
}
\makeatother
\def\mycopyrightnotice{%
  \begin{minipage}{\textwidth}
  \centering \scriptsize
  \copyright 2025 IEEE.  Personal use of this material is permitted.  Permission from IEEE must be obtained for all other uses, in any current or future media, including reprinting/republishing this material for advertising or promotional purposes, creating new collective works, for resale or redistribution to servers or lists, or reuse of any copyrighted component of this work in other works.
  \end{minipage}
}
\makeatother
\begin{document}

\title{\LARGE \bf
Finetuning and Quantization of EEG-Based Foundational BioSignal Models on ECG and PPG Data for Blood Pressure Estimation
}

\author{
Bálint Tóth$^{1}$, Dominik Senti$^{1}$, Thorir Mar Ingolfsson$^{1}$, \\
Jeffrey Zweidler$^{1}$, Alexandre Elsig$^{1}$, Luca Benini$^{1,2}$, Yawei Li$^{1}$
\thanks{$^{1}$ETH Z{\"u}rich, Z{\"u}rich, Switzerland.}
\thanks{$^{2}$DEI, University of Bologna, Bologna, Italy.}
\thanks{Bálint Tóth, Dominik Senti, and Thorir Mar Ingolfsson are co-first authors.}
}

\maketitle
\thispagestyle{IEEEtitlepagestyle}

\begin{abstract}
Blood pressure (BP) is a key indicator of cardiovascular health. As hypertension remains a global cause of morbidity and mortality, accurate, continuous, and non-invasive BP monitoring is therefore of paramount importance. Photoplethysmography  (PPG) and electrocardiography (ECG) can potentially enable continuous BP monitoring, yet training accurate and robust machine learning (ML) models remains challenging due to variability in data quality and patient-specific factors. Recently, multiple research groups explored Electroencephalographic (EEG)--based foundation models and demonstrated their exceptional ability to learn rich temporal resolution. Considering the morphological similarities between different biosignals, the question arises of whether a model pre-trained on one modality can effectively be exploited to improve the accuracy of a different signal type. In this work, we take an initial step towards generalized biosignal foundation models by investigating whether model representations learned from abundant EEG data can effectively be transferred to ECG/PPG data solely with fine-tuning, without the need for large-scale additional pre-training, for the BP estimation task. Evaluations on the MIMIC-III and VitalDB datasets demonstrate that our approach achieves near state-of-the-art accuracy for diastolic BP (mean absolute error of 1.57 mmHg) and surpasses by 1.5$\times$ the accuracy of prior works for systolic BP (mean absolute error 2.72 mmHg). Additionally, we perform dynamic INT8 quantization, reducing the smallest model size by over $3.5\times$ (from $13.73$ MB down to $3.83$ MB) while preserving performance, thereby enabling unobtrusive, real-time BP monitoring on resource-constrained wearable devices.
\newline
\indent \textit{Clinical relevance}—Our cuffless BP estimation framework addresses the limitations of traditional, intermittent cuff-based devices by exploiting common ECG and PPG signals, enabling more frequent and unobtrusive measurement. With EEG-based pre-training and hardware-friendly quantization, this approach holds promise for low-power, continuous BP monitoring in both clinical and home environments, potentially improving early detection of hypertension and related cardiovascular risks.
\end{abstract}


\section{Introduction}
\Gls{bp} is one of the most critical indicators of cardiovascular health and overall physiological status~\cite{vital_signs}. Typically reported as \gls{sbp} and \gls{dbp} pressure, normal adult values range between 90–120~mmHg (\gls{sbp}) and 60–80~mmHg (\gls{dbp})~\cite{bp_range}. Elevated \gls{bp}, or hypertension, is a major risk factor for severe \glspl{cvd}, including stroke and heart failure~\cite{leading_cause,CVDs_num}. In 2017 alone, over 17 million deaths worldwide were linked to CVDs, with high \gls{sbp} implicated in more than half of these cases~\cite{HSBP_num}. These statistics highlight the urgent need for frequent or continuous \gls{bp} monitoring to enable early detection and timely intervention.

Most clinical and at-home \gls{bp} measurements currently rely on cuff-based devices, which are non-invasive but provide only intermittent readings. Their bulkiness and requirement for patient compliance also limit their utility for continuous tracking~\cite{bradley2022cuffless, pilz2024cuff}. Consequently, there is substantial interest in leveraging additional physiological signals—particularly \gls{ecg} and \gls{ppg}—to estimate \gls{bp} without a cuff. \gls{ecg} measures electrical activity of the heart, while \gls{ppg} tracks blood volume changes. Both exhibit strong correlations with arterial blood pressure~\cite{ECG_BOOK,ppg_book}, and recent deep learning methods have successfully modeled these complex relationships~\cite{Faust2018DeepLF,Jamil}.

Despite these advances, several challenges remain. Although large-scale labeled \gls{ecg}/\gls{ppg} datasets exist~\cite{ptb-xl,mimiciii,vitaldb} heterogeneity in data quality and physiological variability across patients can degrade model generalizability. Meanwhile, foundation models have emerged in biosignal research~\cite{CEReBrO,Mehta2023CantTT}, where large neural networks are pre-trained on vast amounts of (often unlabeled) data and then fine-tuned for specific tasks. Importantly, the rich temporal and spectral diversity of EEG signals enables transformer-based encoders to learn robust and generalizable representations that can effectively transferred to other biosignals~\cite{yang2023cross,joshi2021deep}.

In this context, a crucial question arises: \textbf{Can knowledge learned from \gls{eeg} waveforms transfer effectively to other biosignals, such as \gls{ecg} and \gls{ppg}, for tasks such as \gls{bp} estimation?} Although \gls{eeg} signals originate from neural activity, they share core time-series characteristics with \gls{ecg}/\gls{ppg}, potentially enabling cross-biosignal feature transfer~\cite{yang2023cross, ingolfsson2024brainfusenet}. To our knowledge, this work is the first to experimentally validate an \gls{eeg}-based foundation model for cuffless \gls{bp} prediction using knowledge transfer from \gls{eeg} to \gls{ecg} and \gls{ppg} without requiring additional large-scale biosignal-specific pre-trainings. This initial investigation lays the groundwork for future research, wherein a dedicated \gls{ecg}/\gls{ppg} foundation model could be developed by further fine-tuning the pre-trained \gls{eeg} model on extensive \gls{ecg} data.

After demonstrating the \gls{eeg}-to-\gls{ecg}/\gls{ppg} knowledge transfer, a second critical challenge to enable wearability is minimizing the computational requirements. Wearables have strong constraints on computational resources. In this context, quantization techniques are necessary to reduce memory footprint and inference latency~\cite{nagel2021whitepaperneuralnetwork}. Thanks to their ability to compress floating-point weights to lower-precision integer representations, quantization is a key ingredient to complement the development of foundation models for wearables.

In this paper, we demonstrate a cross-biosignal transfer learning approach where a transformer model, pre-trained only on \gls{eeg}, is fine-tuned for \gls{bp} estimation from \gls{ecg} and \gls{ppg} waveforms. In particular, no additional large-scale pre-trainings on \gls{ecg}/\gls{ppg} were required. We extensively evaluate this method on the MIMIC-III and VitalDB datasets, examining the trade-offs of various fine-tuning strategies, including frozen vs.\ unfrozen backbones. Furthermore, we incorporate dynamic INT8 quantization to reduce the model size by over $3.5\times$, enabling feasible deployment on edge devices. 

In summary, the contribution of this paper is as follows:

\begin{itemize}
    \item We show that a large transformer model pre-trained \textit{only} on \gls{eeg} can be successfully adapted to estimate \gls{bp} from \gls{ecg} and \gls{ppg} signals by relying solely on an additional fine-tuning, opening new avenues for cross-biosignal foundation models.

     \item We compare frozen vs.\ unfrozen transformer backbones and training from scratch vs.\ pre-trained weights, quantifying their impact on convergence speed, predictive accuracy, and computational cost.
     
    \item Our empirical results on MIMIC-III and VitalDB achieve near state-of-the-art accuracy for diastolic BP, with a mean absolute error of 1.57 mmHg, and surpass the accuracy of prior works for systolic BP, with a mean absolute error of 2.72 mmHg (1.5 $\times$ smaller than SoA).

    \item We apply dynamic INT8 quantization to reduce the model size by over $3.5\times$ (from 13.73 MB to 3.83 MB)  with negligible performance loss, enabling resource-efficient inference for real-time \gls{bp} monitoring.
\end{itemize}
\section{Related Work}\label{sec:related_works}
\gls{bp} estimation from \gls{ecg} and \gls{ppg} waveforms has received significant attention due to its potential for continuous, unobtrusive monitoring. Earlier work relied on classical machine learning with handcrafted features, but deep learning methods have since emerged as more robust alternatives. Convolutional or recurrent architectures designed for \gls{ecg}/\gls{ppg} have shown strong performance, including ResUNet with self-attention~\cite{Jamil}, U-Net variants~\cite{Mahmud_2022}, and hybrid \gls{cnn}--\gls{rnn} models~\cite{Paviglianiti2021ACO}. These architectures often outperform traditional feature-engineering approaches, particularly when both \gls{ecg} and \gls{ppg} signals are used~\cite{Paviglianiti2021ACO}.

Nevertheless, many existing methods train solely on \gls{ecg}/\gls{ppg} data, which, while plentiful~\cite{mimiciii,vitaldb,ptb-xl}, often exhibit significant variability in signal quality and patient-specific characteristics. This variability poses challenges for achieving robust generalization across populations. Recent work has explored transfer learning to overcome these issues; for example, Yang \emph{et~al.}~\cite{yang2023cross} studied the transfer of \gls{eeg} knowledge to \gls{ecg} classification tasks, achieving improved performance and reduced training costs. Joshi \emph{et~al.}~\cite{joshi2021deep} also explored the transfer of \gls{eeg} knowledge using a deep knowledge distillation framework to enhance single-lead \gls{ecg}-based sleep staging. However, these studies have largely focused on within-modality or narrow domain adaptations, leaving open the broader question of whether an \gls{eeg}-based foundation model can serve as a versatile starting point for generalized biosignal analysis.

\gls{eeg} has become an attractive candidate for pre-training large models not only because of the availability of large-scale \gls{eeg} repositories~\cite{TUEG} but also due to its rich multi-channel, temporal, and spectral dynamics~\cite{jiang2024large}. While many time-series modalities (for example, voice) also exhibit rich temporal structure, \gls{eeg}, \gls{ecg}, and \gls{ppg} share common physiological origins and similar noise characteristics, which facilitate the transfer of temporal pattern recognition capabilities. In other words, our hypothesis is that the underlying statistical properties and multi-dimensional dynamics in \gls{eeg} make it particularly well-suited for learning robust representations that can be effectively adapted to \gls{ecg}/\gls{ppg} tasks. Our work is the first to validate the feasibility of fine-tuning a transformer-based model initially trained on EEG (CEReBrO~\cite{CEReBrO}) for arterial \gls{bp} estimation using \gls{ecg} and \gls{ppg} data.

Beyond accuracy, real-world deployment of \gls{bp} estimation models calls for efficient inference. Traditional deep networks can be computationally expensive, motivating recent interest in quantization and other compression techniques~\cite{nagel2021whitepaperneuralnetwork}. Few studies have combined large-scale pre-training with post-training quantization for \gls{bp} monitoring. Hence, our method integrates these two aspects: leveraging a potent \gls{eeg}-based foundation model and applying quantization for a compact, high-accuracy cuffless \gls{bp} solution.
\section{Methodology}\label{sec:method}

This section describes our approach for estimating \gls{bp} from \gls{ecg} and \gls{ppg} waveforms using a large \gls{eeg}-based foundation model. We first detail how we adapt and fine-tune the CEReBrO architecture for \gls{bp} prediction, then describe our post-training quantization steps.

\subsection{Architecture and Fine-Tuning}\label{subsec:model}

\begin{figure}[htp]
    \centering
    \includegraphics[width=7.5cm]{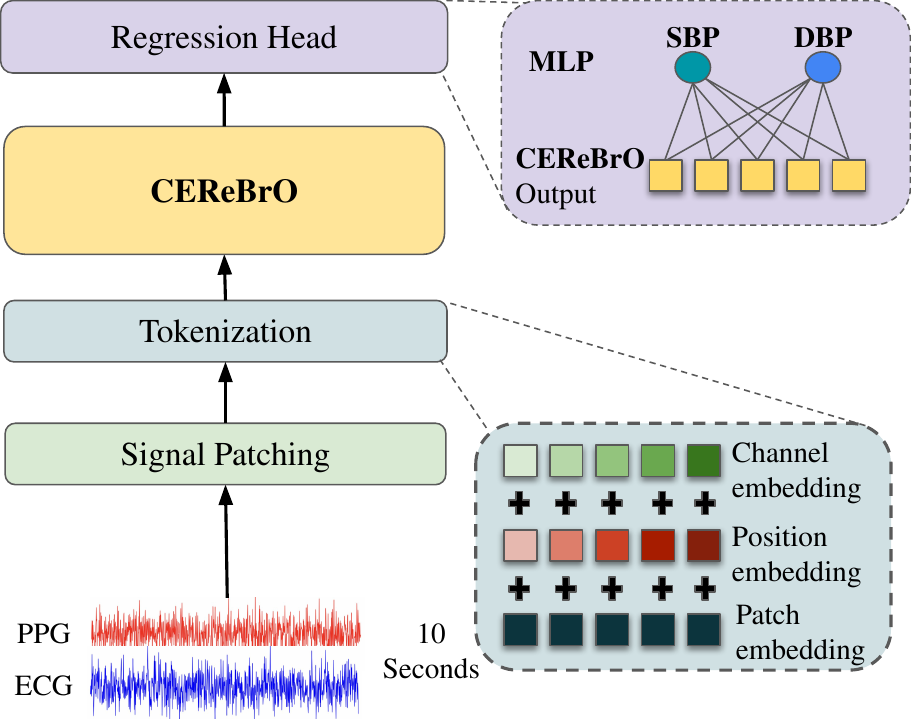}
    \caption{The modified CEReBrO Architecture~\cite{CEReBrO}, supplemented with an additional MLP-head, which is utilized for the \gls{bp} estimation task.} 
    \vspace{-0.6cm}
    \label{fig:cerebro}
\end{figure}  

Our method builds on the \textbf{CEReBrO} transformer-encoder~\cite{CEReBrO}, originally pre-trained on a large \gls{eeg} dataset (TUEG~\cite{TUEG}). CEReBrO employs a tokenization scheme that splits time-series signals into non-overlapping patches and projects them into a latent space. Alternating self-attention blocks then process these tokens by focusing on intra-channel (temporal) correlations and inter-channel (spatial) relationships. This design efficiently captures both local and long-range dependencies in multi-channel biosignals. Although CEReBrO was trained on \gls{eeg} data, its attention-based encoder can generalize to other biosignals sharing similar temporal structures. To adapt CEReBrO for \gls{bp} estimation from \gls{ecg} and \gls{ppg}, we make the following modifications:

\begin{itemize}
    \item We feed \gls{ecg} and \gls{ppg} signals as two input channels, each sampled at 125,Hz and shaped into 10-second segments ($2 \times 1250$).
    \item  We replace the original classification head with a single fully connected layer (MLP) that outputs two values: \gls{sbp} and \gls{dbp}.
\end{itemize}

CEReBrO is then also available in three sizes—\emph{small} (3.58M parameters), \emph{medium} (39.95M parameters), and \emph{large} (85.15M parameters). During fine-tuning, we largely retained the original hyperparameters from the pre-trained CEREBRO model. We set the learning rate to 1e-4 and used a batch size of 1024, selected primarily to optimize GPU memory utilization rather than through an extensive hyperparameter search.

We then explore two fine-tuning strategies:
\begin{itemize}
    \item \textbf{Frozen Backbone}: All transformer layers except for the first input-embedding layer and the final MLP head are frozen. This preserves most \gls{eeg}-based representations while allowing partial adaptation to \gls{ecg}/\gls{ppg}.
    \item \textbf{Unfrozen Backbone}: All transformer layers are unfrozen to allow deeper domain alignment, albeit with a risk of forgetting learned \gls{eeg} features.
\end{itemize}
To measure the benefit of leveraging a pre-trained \gls{eeg} encoder, we compare fine-tuning against training from scratch (i.e., random initialization). In each setting, we train models of three different sizes (small, medium, large) for 100 epochs and extend unfrozen-backbone runs to 200 epochs to assess longer-term convergence. We use Xavier Initialization~\cite{xavier_init} when training from scratch. Performance is evaluated on the MIMIC-III and VitalDB datasets in terms of MAE, SD, and coefficient of determination ($R^2$), as well as clinical standards (\gls{bhs} and \gls{aami}). To mitigate the risk of overfitting during the fine-tuning of these large models, we employed standard validation techniques. Data was partitioned into training, validation, and test sets, and we utilized an early stopping strategy based on performance monitoring on the validation set to select the best model checkpoint and prevent over-specialization to the training data
Note that while the foundation model was pre-trained on EEG data, our BP estimation workflow uses only ECG and PPG signals during fine-tuning and inference; no patient-specific EEG calibration is required.

\subsection{Quantization}\label{subsec:quantization}
We apply post-training quantization to our fine-tuned models to enable real-time deployment on resource-constrained devices. This step reduces the memory footprint and inference latency while preserving clinically relevant accuracy.

We use PyTorch’s FX Graph Mode Quantization pipeline~\cite{pytorch2} to insert quantization and dequantization operations systematically. Quantization is widely employed to map floating-point (32-bit) values to lower numerical precision, typically to 8-bit integers. The range of a floating-point value, denoted by \(x_{\mathrm{fp}}\), is defined as follows: \([x_{\min}, x_{\max}]\). Based on this, two characteristic values can be defined, which are essential for the quantization process: \textbf{Scale} \(\Delta\), which determines the step size (real-valued), while \textbf{Zero-point} \(z\), which is an integer offset whose primary function is to ensure that the zero is mapped onto an integer. In this work, we specifically adopt two types of quantization:

\begin{itemize}
    \item \emph{Symmetric Quantization} for weights (common when weight distributions are roughly zero-mean).
    \item \emph{Asymmetric Quantization} for activations (typical when ReLU shifts values positively).
\end{itemize}
For symmetric quantization, we have the following characteristic values:
\begin{equation*}
    \Delta = \frac{\max\bigl(|x_{\min}|, |x_{\max}|\bigr)}{2^{b-1}}, 
    \quad 
    z = 0
\end{equation*}

Based on these values, forward quantization is done using this equation:
\begin{equation*}
x_{\mathrm{int}} = \mathrm{clip}\!\Bigl(
        \mathrm{round}\bigl(\tfrac{x}{\Delta}\bigr), 
        \, -2^{b-1}, \, 2^{b-1} - 1\Bigr).
\end{equation*}

And for \emph{asymmetric} quantization the characteristic values can be calculated in the following way, where $b$ is the number of bits :
\begin{equation*}
    \Delta = \frac{x_{\max} - x_{\min}}{2^b - 1}, 
    \quad 
    z = \left\lfloor - \frac{x_{\min}}{\Delta} + 0.5 \right\rfloor.
\end{equation*}
When $\Delta$ and $z$ are determined, the forward quantization step is the following:
\begin{equation*}
    x_{\mathrm{int}} = \mathrm{clip}\!\Bigl(
        \mathrm{round}\bigl(\tfrac{x_{\mathrm{fp}}}{\Delta}\bigr) + z , 
        \, 0, \, 2^b - 1\Bigr),
\end{equation*}
where $\mathrm{clip}(\cdot,0,2^b-1)$ ensures $x_{\mathrm{int}}$ to stay in the range $[0, 2^b - 1]$~\cite{quant_data}.

Quantization typically involves three stages: (1) \emph{calibration}, where representative data is passed through the model to collect scaling statistics; (2) \emph{conversion}, transforming the floating-point model into a quantized version; and (3) \emph{execution}, running inference with reduced-precision operations.

We explore both \emph{static} quantization~\cite{FU20092937}, which precomputes scaling and zero points via a calibration dataset, and \emph{dynamic} quantization~\cite{vu2008stabilizing}, which calculates them on-the-fly, eliminating the calibration phase. While static quantization can offer speed benefits if the input distribution is stable, dynamic quantization is often more flexible for variable-length or varying data distributions and avoids the need for extra calibration data.

Our target precision is INT8, balancing memory savings and model fidelity. We evaluate symmetric quantization for weights and asymmetric for activations (shifted by ReLU). Different observers—\emph{MinMaxObserver}, \emph{MovingAverageMinMaxObserver}, and \emph{HistogramObserver}—estimate the range, each trading off complexity against robustness. We also employ per-channel quantization for \gls{ecg}/\gls{ppg} inputs, giving each signal channel a separate scale and zero point.

Our experiments reveal that dynamic per-channel quantization with symmetric weights yields an optimal model size, computational speed, and accuracy trade-off. Detailed results of these experiments are presented in Section~\ref{sec:results}. This approach is critical for enabling continuous, on-device \gls{bp} estimation, where both memory and energy constraints are strict.
\section{Experimental Results}\label{sec:results}
This section presents a comprehensive evaluation of our approach on two widely used public datasets, \textbf{MIMIC-III} and \textbf{VitalDB}, accessed through the pre-processed \textbf{PulseDB} resource~\cite{pulsedb}. After describing these datasets and their integration in PulseDB, we introduce the evaluation metrics used for both predictive and clinical quality. We then discuss the fine-tuning experiments, focusing on how frozen/unfrozen backbones and extended epochs influence performance. Finally, we assess the impact of quantization, highlighting both accuracy and practical gains.

\begin{table}[b]
\vspace{-0.5cm}
    \footnotesize
    \centering
    \caption{BHS Grading System~\cite{BHS}}\label{tab:req}
    \begin{tabular}{l c c c} 
        \toprule
        Grades & $\leq 5 \, \text{mmHg}$ & $\leq 10 \, \text{mmHg}$ & $\leq 15 \, \text{mmHg}$ \\
        \midrule
        Grade A & 60\% & 85\% & 95\% \\
        Grade B & 50\% & 75\% & 90\% \\
        Grade C & 40\% & 65\% & 85\% \\ 
        \bottomrule
    \end{tabular}
\end{table}

\subsection{Datasets and Preprocessing}\label{sec:datasets}
\textbf{PulseDB} aggregates and cleans segments derived from both MIMIC-III and VitalDB. It provides 10-second windows of synchronized \gls{ecg}, \gls{ppg}, and \gls{abp} signals, downsampled to 125\,Hz. A total of 4,941 records are extracted from MIMIC-III and 3,458 from VitalDB~\cite{pulsedb}. By leveraging PulseDB, we work with consistent data splits and standardized preprocessing.

\begin{itemize}
    \item \textbf{MIMIC-III}~\cite{mimiciii}, collected from around 30,000 intensive care unit (ICU) patients MIMIC-III includes time-series of \gls{ecg}, \gls{ppg}, and \gls{abp} measurements at 125\,Hz. PulseDB extracts high-quality 10-second windows to form our training and evaluation splits.
    \item \textbf{VitalDB}~\cite{vitaldb} contains synchronized recordings from 6,153 subjects, originally sampled at 500\,Hz. We down-sampled these signals to 125\,Hz to match MIMIC-III, ensuring uniform sampling rates. 
\end{itemize}

Unless otherwise noted, each dataset (MIMIC-III portion and VitalDB portion within PulseDB) is randomly divided into 80\% training, 10\% validation, and 10\% testing. This setup allows us to compare performance across two distinct clinical data sources under consistent preprocessing.

\subsection{Evaluation Metrics and Clinical Standards}\label{subsec:metrics}

We employ three primary statistical metrics to assess the predictive accuracy of our BP estimation models, along with two clinical standards that gauge practical viability. Let $\text{Error} = BP_{\text{target}} - BP_{\text{estimated}}$ be the difference between the ground-truth BP and the model prediction. We report the MAE, which highlights how significant the prediction errors are on average. SD, which captures the spread or consistency of the errors. And the coefficient of determination ($R^2$) measures how well the model fits the observed data. These metrics reveal both the average error magnitude and the overall fit.

\begin{itemize}
    \item \textit{\gls{bhs}:} Graded A--D based on the percentage of predictions falling within $5$, $10$, and $15$\,mmHg of ground truth~\cite{BHS}. Grade~A requires at least 60\% of errors $\le5$\,mmHg, 85\% $\le10$\,mmHg, and 95\% $\le15$\,mmHg.
    \item \textit{\gls{aami}:} Considers a model valid if its mean bias is within $\pm5$\,mmHg and the SD of errors is below 8\,mmHg~\cite{AAMI}.
\end{itemize}

\begin{table*}[ht!]
\centering
\caption{Combined Results: Fine-Tuning (FT) vs.\ Training-from-Scratch (TFS), 100 vs.\ 200 Epochs, Frozen vs.\ Unfrozen, Across MIMIC-III and VitalDB. \gls{bhs} and \gls{aami} Columns Indicate Clinical Standard Compliance.}
\label{tab:merged_100_200}
\resizebox{\textwidth}{!}{
\begin{tabular}{llccccccccccl}
\toprule
\textbf{Dataset} & \textbf{Method} & \textbf{Frozen?} & \textbf{Epochs} & \textbf{Size} & \textbf{DBP SD} & \textbf{DBP MAE} & 
\textbf{DBP R$^2$} & \textbf{SBP SD} & \textbf{SBP MAE} & \textbf{SBP R$^2$} & \textbf{BHS} & \textbf{AAMI}
\\
\midrule
MIMIC & FT & False & 100 & Small &
4.12 & 2.58 & 0.90 & 5.97 & 4.13 & 0.93 & A & Pass \\
MIMIC & FT & False & 100 & Medium &
3.57 & 2.05 & 0.93 & 5.19 & 3.42 & 0.95 & A & Pass \\
MIMIC & FT & False & 100 & Large &
3.18 & 1.72 & 0.94 & 4.55 & 2.93 & 0.96 & A & Pass \\
\cmidrule(lr){2-13}
MIMIC & FT & False & 200 & Small &
3.63 & 2.19 & 0.93 & 5.26 & 3.58 & 0.95 & A & Pass \\
MIMIC & FT & False & 200 & Medium &
3.29 & 1.80 & 0.94 & 4.78 & 3.05 & 0.96 & A & Pass \\
MIMIC & FT & False & 200 & Large &
\textbf{3.04} & \textbf{1.57} & \textbf{0.95} & \textbf{4.35} & \textbf{2.72} & \textbf{0.96} & A & Pass \\
\cmidrule(lr){2-13}
MIMIC & TFS & False & 100 & Small &
4.55 & 2.89 & 0.88 & 6.69 & 4.67 & 0.91 & A & Pass \\
MIMIC & TFS & False & 100 & Medium &
3.96 & 2.32 & 0.91 & 5.87 & 3.91 & 0.93 & A & Pass \\
MIMIC & TFS & False & 100 & Large &
3.83 & 2.23 & 0.92 & 5.66 & 3.71 & 0.94 & A & Pass \\
\cmidrule(lr){2-13}
MIMIC & FT & True & 100 & Small &
13.13 & 10.11 & 0.03 & 22.29 & 17.92 & 0.04 & D & Fail \\
\midrule
VitalDB & FT & False & 100 & Small &
6.63 & 4.77 & 0.70 & 9.70 & 6.99 & 0.74 & C & Pass/Fail \\
VitalDB & FT & False & 100 & Medium &
5.49 & 3.67 & 0.79 & 8.19 & 5.56 & 0.81 & A & Pass/Fail \\
VitalDB & FT & False & 100 & Large &
3.90 & 2.39 & 0.90 & 5.85 & 3.77 & 0.91 & A & Pass \\
\cmidrule(lr){2-13}
VitalDB & FT & False & 200 & Large &
\textbf{3.42} & \textbf{1.92} & \textbf{0.92} & \textbf{5.12} & \textbf{3.14} & \textbf{0.93} & A & Pass \\
\cmidrule(lr){2-13}
VitalDB & TFS & False & 100 & Medium &
6.14 & 4.18 & 0.74 & 9.25 & 6.39 & 0.76 & A & Pass/Fail \\
VitalDB & TFS & False & 100 & Large &
3.90 & 2.39 & 0.90 & 5.85 & 3.77 & 0.91 & A & Pass \\
\cmidrule(lr){2-13}
VitalDB & FT & True & 100 & Large &
11.45 & 9.26 & 0.08 & 18.20 & 14.78 & 0.06 & D & Fail \\
\bottomrule
\end{tabular}
}
\vspace{-0.5cm}
\end{table*}

\subsection{Fine-tuning Experiments}\label{subsec:finetune_results}

We evaluated our fine-tuning approach under varying settings (1):(1) \textit{Frozen vs.\ Unfrozen} backbones, (2) \textit{Small/Medium/Large} model sizes, and (3) \textit{100 vs.\ 200} training epochs. For comparison, we also trained all model sizes from scratch to quantify the benefit of leveraging a pre-trained \gls{eeg} foundation. 

Table~\ref{tab:merged_100_200} \emph{consolidates} all key results for experiments utilizing 100 and 200 epochs. As shown, across all three model sizes (S, M, L), using an \emph{unfrozen} backbone consistently yields better MAE and SD values than freezing it. On MIMIC-III, the large (L) model achieves the lowest MAE (1.72~mmHg for \gls{dbp} and 2.93~mmHg for \gls{sbp}) and highest $R^2$ (0.94 for \gls{dbp}, 0.96 for \gls{sbp}) at 100 epochs. On VitalDB, fine-tuning likewise improves accuracy, with the large model reaching an MAE of 2.39~mmHg for \gls{dbp} and 3.77~mmHg for \gls{sbp} at 100 epochs. Notably, these models often satisfy \gls{aami} standards and achieve \gls{bhs} Grade~A or B.

By contrast, models trained from scratch require more epochs to converge and exhibit higher MAE and SD. For instance, on MIMIC-III, the large model trained from scratch attains an MAE of 2.23~mmHg (\gls{dbp}) and 3.71~mmHg (\gls{sbp}), compared to 1.72~mmHg and 2.93~mmHg for the fine-tuned version—demonstrating how leveraging a pre-trained backbone reduces error and speeds convergence.

To examine whether additional training refines the model further, we extended the unfrozen fine-tuning runs to 200 epochs. The medium and large models on MIMIC-III increase $R^2$ above 0.94 for \gls{dbp} and 0.96 for \gls{sbp}, corroborating the advantages of longer training. On VitalDB, the large model’s MAE drops to 1.92~mmHg (\gls{dbp}) and 3.14~mmHg (\gls{sbp}), also improving $R^2$ to 0.92 and 0.93, respectively. Depending on clinical requirements and computational budget, these incremental gains may be justified—particularly in offline training where slight improvements can enhance overall reliability.

Table~\ref{tab:merged_100_200} displays the \gls{bhs} grades and \gls{aami} standard compliance for all 100-epoch runs. Fine-tuning typically achieves \gls{bhs} Grade~A or B for both \gls{dbp} and \gls{sbp}, and meets the \gls{aami} criteria of $\pm5$~mmHg mean error and $<8$~mmHg SD. Notably, the large CEReBrO model meets Grade~A for VitalDB in both \gls{dbp} and \gls{sbp} under unfrozen fine-tuning, indicating near-clinical reliability.

Overall, these results demonstrate that \emph{unfrozen fine-tuning of a pre-trained \gls{eeg} foundation model} yields the best trade-off among accuracy, robustness, and training efficiency across two significant datasets. More importantly, these findings confirm that a pre-trained \gls{eeg} transformer can robustly adapt to \gls{ecg}/\gls{ppg}-based BP estimation, reducing errors while meeting rigorous clinical standards.

\subsection{Quantization Performance}\label{subsec:quant_restults}
We evaluate static vs.\ dynamic INT8 quantization (Section~\ref{subsec:quantization}) on our best-fine-tuned models (unfrozen, 200 epochs), examining the size-accuracy trade-off.

As shown in Table~\ref{tab:quantization_r2}, \emph{dynamic per-channel quantization with symmetric weights} yields the highest $R^2$ for both \gls{dbp} and \gls{sbp} across MIMIC-III and VitalDB, closely matching the unquantized baseline. In particular, the large model’s \gls{dbp} $R^2$ only drops from 0.9479 to 0.9476 on MIMIC-III, a negligible difference in performance.

Although static quantization yields similar compression ratios, Table~\ref{tab:model_size_comparison} reveals a negligible difference in final model size compared to dynamic quantization while exhibiting a slightly lower $R^2$ in practice, which is shown in Table~\ref{tab:quantization_r2}. Additionally, static quantization requires a separate calibration step that can complicate deployment.

Table~\ref{tab:model_size_comparison} highlights a reduction factor of approximately 3.5--3.9$\times$ across all model sizes. The smallest model shrinks to 3.85~MB, enabling truly resource-limited scenarios. Meanwhile, the large model is compressed to around 83~MB yet preserves top-tier predictive performance while also being suitable for embedded deployment on devices with suitable flash sizes. The quantized models also meet \gls{aami} standards and achieve \gls{bhs} Grade~A on both datasets, indicating minimal performance degradation compared to the unquantized baseline.

\begin{table}[b]
\centering
\vspace{-0.5cm}
\caption{Dynamic and Static Quantization test $R^2$ versus original test $R^2$ for DBP and SBP on the 200 epoch models}
\label{tab:quantization_r2}
\renewcommand{\arraystretch}{1.2} 
\setlength{\tabcolsep}{2pt} 
\resizebox{\columnwidth}{!}{
\begin{tabular}{llcccccc}
\toprule
\textbf{Dataset} & \textbf{Size} & \multicolumn{2}{c}{\textbf{Original}} & \multicolumn{2}{c}{\textbf{Dynamic Quantized}} & \multicolumn{2}{c}{\textbf{Static Quantized}} \\
\cmidrule(lr){3-4} \cmidrule(lr){5-6} \cmidrule(lr){7-8}
& & \textbf{DBP R$^2$} & \textbf{SBP R$^2$} & \textbf{DBP R$^2$} & \textbf{SBP R$^2$} & \textbf{DBP R$^2$} & \textbf{SBP R$^2$} \\
\midrule
\multirow{3}{*}{MIMIC} & Small & 0.9256 & 0.9467 & 0.9253 & 0.9465 & 0.9247 & 0.9460 \\
& Medium & 0.9389 & 0.9559 & 0.9386 & 0.9556 & 0.9374 & 0.9546 \\
& Large & 0.9479 & 0.9635 & 0.9476 & 0.9632 & 0.9466 & 0.9628 \\
\midrule
\multirow{3}{*}{VitalDB} & Small & 0.8034 & 0.8304 & 0.8019 & 0.8295 & 0.7994 & 0.8279 \\
& Medium & 0.8534 & 0.8659 & 0.8522 & 0.8652 & 0.8503 & 0.8634 \\
& Large & 0.9203 & 0.9280 & 0.9198 & 0.9276 & 0.9184 & 0.9265 \\
\bottomrule
\end{tabular}
}
\end{table}

\begin{table}[ht!]
\centering
\caption{Comparison of model sizes for the original, dynamically quantized, and statically quantized models across different model sizes (measured in MB). RF stands for reduction factor.}
\vspace{-0.1cm}
\label{tab:model_size_comparison}
\begin{tabular}{lcccc}
\toprule
\textbf{Size} & \textbf{Original (MB)} & \textbf{Dynamic (MB)} & \textbf{Static (MB)} & \textbf{RF} \\
\midrule
Small & 13.73 & 3.85 & 3.83 & 3.57 \\
Medium & 152.55 & 39.47 & 39.43 & 3.87 \\
Large & 324.96 & 83.28 & 83.22 & 3.90 \\
\bottomrule
\end{tabular}

\vspace{-0.5cm}
\end{table}

\subsection{Comparison to SOTA}
Table~\ref{tab:comparison_combined} compares our work to \gls{soa}. Our proposed method demonstrates a clear performance advantage over existing state-of-the-art approaches on both the VitalDB and MIMIC-III datasets. On the VitalDB dataset, our model achieves a DBP MAE of 1.92 mmHg and an SBP MAE of 3.14 mmHg. These results represent a reduction of approximately 35\% in DBP error and 32\% in SBP error compared to the ResUNet+Attention approach by Jamil et al.~\cite{Jamil}, which reported MAEs of 2.95 mmHg and 4.64 mmHg for DBP and SBP respectively. Furthermore, our model outperforms other recent methods such as CiGNN~\cite{liu2024cignn}, rU-Net~\cite{runet}, and Fusion+Attention~\cite{el2021deep}, where the latter reported even higher errors (3.76 mmHg for DBP and 5.32 mmHg for SBP).
Our method achieves a DBP MAE of 1.57 mmHg and an SBP MAE of 2.72 mmHg on the MIMIC-III dataset. Notably, while the ResUNet+Attention method by Jamil et al.~\cite{Jamil} produced a slightly lower DBP error of 1.13 mmHg, its SBP error was significantly higher at 4.55 mmHg, making our SBP performance nearly 40\% better. Moreover, when compared to the hybrid ResNet-LSTM approach by Paviglianiti et al.~\cite{Paviglianiti2021ACO}, which reported errors of 2.23 mmHg (DBP) and 4.12 mmHg (SBP), our method demonstrates improvements of approximately 29\% and 34\% for DBP and SBP respectively. The performance gains over additional methods such as GWO-GBRT~\cite{liu2024continuous} and Conv-LSTM~\cite{kamanditya2024continuous} further underscore the robustness of our approach. These quantitative improvements confirm that our EEG-based Foundation BioSignal design not only rivals but, in many respects, surpasses the current state-of-the-art while also offering enhanced scalability and feasibility for edge deployment. 

\begin{table}[b]
\vspace{-0.5cm}
\centering
\caption{Comparison of MAE across VitalDB and MIMIC-III datasets (all values in mmHg)}
\label{tab:comparison_combined}
\resizebox{\columnwidth}{!}{
\begin{tabular}{llcc}
\toprule
\textbf{Dataset} & \textbf{Method} & \textbf{DBP MAE} & \textbf{SBP MAE} \\
\midrule
\multirow{4}{*}{VitalDB} & ResUNet+Attention~\cite{Jamil} & 2.95 & 4.64 \\
& CiGNN~\cite{liu2024cignn} & 2.79 & 4.15 \\
& Fusion+Attention~\cite{el2021deep} & 3.76 & 5.32 \\
& rU-Net~\cite{runet} & 2.69 & 4.49 \\
& \textbf{This Work} & \textbf{1.92} & \textbf{3.14} \\
\midrule
\multirow{4}{*}{MIMIC-III} & ResUNet+Attention~\cite{Jamil} & \textbf{1.13} & 4.55 \\
& GWO-GBRT~\cite{liu2024continuous} & 2.79 & 4.15 \\
& Conv-LSTM~\cite{kamanditya2024continuous} & 3.29 & 4.30 \\
& ResNet+LSTM~\cite{Paviglianiti2021ACO} & 2.23 & 4.12 \\
& \textbf{This Work} & 1.57 & \textbf{2.72} \\
\bottomrule
\end{tabular}
}
\end{table}

\subsection{Discussion}
Our findings confirm the hypothesis that it is possible to leverage pre-trained \gls{eeg} representations to enable accurate, cuffless blood pressure monitoring from \gls{ecg} and \gls{ppg} signals. By fine-tuning a large transformer-based backbone originally trained on \gls{eeg}, we achieve low error rates that align with clinical benchmarks such as the \gls{bhs} grading system and \gls{aami} standards. Despite these encouraging results, several areas warrant deeper exploration and refinement.

First, data heterogeneity remains a significant concern. Although we validated our approach on MIMIC-III and VitalDB, these datasets exhibit variations in patient demographics, data quality, and clinical conditions that may not fully capture the breadth of real-world scenarios. Extending the model to broader populations or leveraging domain-adaptation techniques could bolster robustness across diverse settings. Moreover, while our \gls{eeg}-to-\gls{ecg}/\gls{ppg} transfer highlights that time-series features learned from brain activity can generalize to cardiovascular signals, a more granular examination of which latent representations transfer most effectively—and whether additional modalities might further enhance performance—could inform future multi-modal foundation models.

Second, our work demonstrates that quantization cuts model size and computational demands with minimal accuracy loss, thereby opening a viable path for embedded or wearable devices. However, real-time inference on resource-constrained hardware introduces additional factors such as on-device latency, battery constraints, and environmental variability. Evaluating our quantized model on ultra-low-power microcontrollers remains essential for future developments towards fully wearable deployments. Techniques such as structured pruning or model distillation may further optimize the trade-off between model capacity and energy consumption.

Finally, although we meet clinical standards under controlled conditions, truly continuous monitoring in daily life entails challenges like motion artifacts, intermittent signal dropouts, and inconsistent sensor placements. Advanced artifact minimization strategies~\cite{ingolfsson_minimizing_2024} and sensor fusion with inertial measurement units~\cite{ingolfsson2024brainfusenet} may mitigate these real-world confounders. In tandem with demographic or patient-specific calibration, such approaches could help ensure that cuffless BP estimation remains reliable over prolonged usage. These directions illuminate how cross-biosignal transfer, combined with robust compression techniques, can evolve into a clinically meaningful framework for unobtrusive, continuous blood pressure monitoring.

\section{Conclusion}
In this work, we demonstrated that a large transformer-based foundation model, originally trained on EEG, can be successfully adapted to estimate blood pressure from ECG and PPG with near-clinical accuracy. Using the CEReBrO model~\cite{CEReBrO} as our base, we achieved MAE as low as 1.57~mmHg (DBP) and 2.72~mmHg (SBP) on the MIMIC-III dataset, and 1.92~mmHg (DBP) and 3.14~mmHg (SBP) on VitalDB. These results fulfill critical benchmarks such as the BHS Grade~A and the AAMI guidelines, reinforcing the clinical viability of cross-biosignal knowledge transfer.

Beyond raw accuracy, we showed that post-training quantization substantially reduces the model footprint with minimal performance degradation. Dynamic quantization compressed the largest model from over $300$~MB to approximately $80$~MB—an almost $3.8\times$ reduction—while preserving MAE and \(R^2\) scores. This compression is vital for on-device, real-time BP monitoring in energy-constrained wearable systems, enabling more frequent and unobtrusive cardiovascular health assessments compared to traditional cuff-based devices.

Our findings underscore the promise of foundation models in biosignal processing and illustrate how quantization fosters practical deployment in low-resource or portable settings. Importantly, this work lays the groundwork for future research wherein the EEG-based foundation model not only serves as an effective tool for cross-biosignal transfer but also as a basis for developing dedicated ECG/PPG foundation models. Future work will explore additional compression methods such as pruning and distillation, integration of demographic factors for improved fairness, and further validation on embedded platforms to handle real-world artifacts and motion conditions. By combining efficient cross-biosignal transfer and hardware-friendly optimizations, this approach offers a path toward continuous, real-time blood pressure monitoring that can enhance patient care and clinical decision-making.
\vspace{-0.15cm}
\section*{Acknowledgment}
\vspace{-0.15cm}
This project is supported by the Swiss National Science Foundation under the grant number 193813 (PEDESITE project) and by the ETH Future Computing Laboratory (EFCL), financed by a donation from Huawei. This work was supported by a grant from the Swiss National Supercomputing Centre (CSCS) under project ID lp12 on Alps.
\bibliographystyle{IEEEtran}
\bibliography{citations}
\end{document}